\documentclass[aps,prb,twocolumn,superscriptaddress,nofootinbib,longbibliography]{revtex4-1}
\usepackage{epsfig,amsopn}
\usepackage{graphicx}
\usepackage{color}
\usepackage{amsmath,amssymb}
\usepackage{enumerate}
\newcommand\bea{\begin{eqnarray}}
\newcommand\eea{\end{eqnarray}}
\newcommand\beq{\begin{equation}}
\newcommand\eeq{\end{equation}}

\def\nn{\nonumber}
\def\f{\frac}

\def\ep{\epsilon}

\def\si{\sigma}

\def\De{\Delta}
\def\dg{\dagger}

\def\ua{\uparrow}
\def\da{\downarrow}

\begin{document}
\title{DC Josephson effect in superconductor-quantum dot-superconductor junctions} 
\author{ Abhiram Soori~}
\email{abhirams@uohyd.ac.in}
\affiliation{ School of Physics, University of Hyderabad, Gachibowli, Hyderabad 500046, India.}

\begin{abstract}
A quantum dot weakly coupled to two normal metal leads exhibits resonant transmission
when one of the dot energy levels lies within the applied bias window. But when the 
quantum dot is sidecoupled to the transport channel, transmission in the channel is
suppressed when a dot energy lies in the bias window. A steady current can also be
driven in a transport channel by connecting it to superconducting reservoirs and
applying a Josephson phase difference instead of a voltage bias. An interesting
question is to investigate the transport across quantum dot connected to two 
superconductors maintained at a superconducting phase difference. To incorporate
the geometry where quantum dot is sidecoupled, we consider a quantum dot with two
sites connected to the superconductors in two geometrical configurations: (A)~the one 
where both the sites are in the transport channel and (B)~the other where only one 
site is in the transport channel and the second site sidecoupled. We find that both
the configurations show resonant transmission for Josephson current and give 
qualitatively same result when the onsite energies of the two sites in the dot are
equal. The two configurations exhibit distinct Josephson current characteristics when the 
onsite energies of the two sites are equal in magnitude and opposite in sign. We 
understand the obtained results. The systems studied are within the reach of current 
experiments.   
\end{abstract}
%\pacs{}
\maketitle
 
\section{Introduction}
In 1962, Josephson made the phenomenal prediction that across two superconductors
maintained at different phases separated by a thin insulating layer, a DC current 
proportional to the sine of the phase difference should pass~\cite{josephson62}. 
The Josephson current between two superconductors is carried by Andreev bound states
whose energy lies within the superconducting gap~\cite{furusaki99}. 
Josephson junctions are important for various technological reasons. For example, 
SQUIDs designed on the principle of Josephson effect can measure very low magnetic
fields~\cite{drung07}. Also, Josephson effect has been very useful in the fundamental 
research. For example, the periodicity in Josephson effect is used to distinguish 
topological superconductors from nontopological 
superconductors~\cite{rokhi2012,Laroche2019}. Further, quantum bits synthesized from 
Josephson junctions hold promise in the development of 
quantum computers~~\cite{wendin07,Clarke2008}. The Josephson junction in 
point contact geometry can be modeled by a one dimensional continuum (lattice) 
model with a tunnel~barrier~(hopping) between the two superconductors. A current 
can be driven in a channel across a barrier by the application of a bias. Also, 
current can be driven by the application of time dependent potentials in the 
channel~\cite{thouless83,switkes99,soori10}. Quantum dots are tiny mesoscale 
systems where a current can be driven by the application of a 
bias~\cite{vanWees89,soori2009}. But driving a current across a quantum dot  in a 
system where a quantum dot is coupled to two superconducting reservoirs maintained 
at two different phases is relatively less studied~\cite{jorgensen2007,beenakker1992}.
Recently we studied the interplay between pumping and Josephson effect in systems where 
quantum dots are connected to superconductors on either sides~\cite{soori2020}, but a 
study of effects of transmission resonance in such systems is missing. 
Motivated by these facts, we undertake the study of a simple noninteracting quantum dot 
coupled to two superconductors maintained at a phase difference.
Though, Coulomb interactions are important generally in quantum dots with both  
spins~\cite{meden19}, a carbon nanotube quantum dot where interactions can 
be neglected have been coupled to superconducting leads experimentally~\cite{herrero06},
making noninteracting quantum dots experimentally accessible.

In this work, we study different configurations of quantum dot consisting of two sites of
each spin connected to two superconductors.  We model the  superconductors on a finite
lattice  connected to the quantum dot sites and diagonalize the entire Hamiltonian 
numerically. The reason for considering finite systems is twofold. The modes that 
contribute to the Josephson current decay into the superconductors away from the
junction making a finitely many sites near the junction important in the dynamics. 
Another reason why we consider finitely many sites is that the problem becomes exactly
diagonalizable at a reasonable numerical cost. 
From the eigenstates, we calculate the Josephson current. The dot can be 
arranged so that (i)~both the sites are in the transport channel~(configuration~A) 
or~(ii)~only one site of the dot is in the transport channel coupled to both 
the reservoirs, while the second site of the dot is coupled to just the first site 
of the dot~(configuration~B) as shown in Fig.~\ref{fig-schematic}. 
We first study the dependence of Josephson current on superconducting phase difference~$\phi$ and onsite 
energy~$\ep_d$ on both sites of the dot for the two configurations. Since the two configurations give 
qualitatively similar results, we further tweak the two configurations by applying a differential 
onsite energies~$\pm\ep_d$ on the two sites. This tweaking shows a qualitative difference in the results
for the two configurations. These results on Josephson currents in different configurations
are then compared with the zero bias conductance of the dot versus the dot onsite energy
$\ep_d$ in order to get complete understanding of the results.

The paper is organized as follows. The section ``Model and Calculations'' describes the models for the two
configurations and outlines the calculations performed. The following section presents the results and the analysis. 
This is followed by a section that puts together a summary of the work. 
%% Cite~\cite{titov06,das08,sriram19}.
\section{Model and Calculations}
\begin{figure*}
 \includegraphics[scale=1.0]{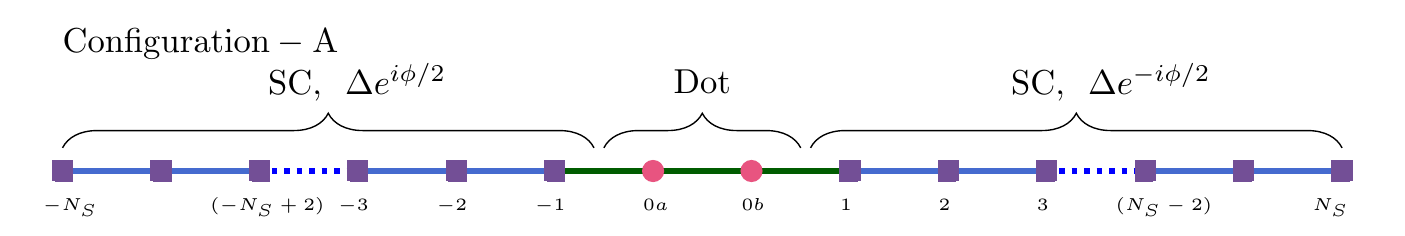}
 \includegraphics[scale=1.0]{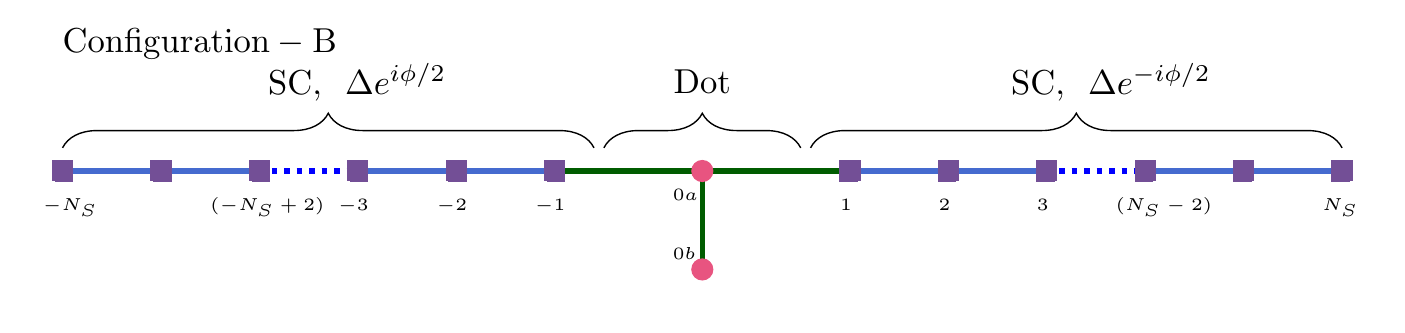}
 \caption{Schematic diagram of the two configurations of the system studied. The quantum dot comprising of sites
 labeled $0a$ and $0b$ is coupled to two superconductors in two configurations. Each superconductor has $N_S$ sites. 
 The superconductor on the left~(right) has a phase $e^{i\phi/2}$($e^{-i\phi/2}$). See Eq.~\eqref{eq:ham} for the full
 Hamiltonians.}~\label{fig-schematic}
\end{figure*}
We study two configurations of the system shown in Fig.~\ref{fig-schematic}. The Hamiltonians $H_{A/B}$ for
configurations~$A/B$ can be written as:
\bea
H_A &=& H_L+H_R+H_D+H_{TA},\nn \\
H_B &=& H_L+H_R+H_D+H_{TB},\nn \\
&&{\rm where~~}\nn \\ 
H_L&=&-t\sum_{n=-1}^{-N_S+1}(c^{\dg}_{n-1}\tau_zc_{n}+{\rm h.c.})\nn\\ &&+\sum_{n=-1}^{-N_S}c^{\dg}_{n}\big[-\mu \tau_z
+\De\cos{(\f{\phi}{2})}\tau_x+\De\sin{(\f{\phi}{2})}\tau_y \big]c_{n}, \nn \\
H_R&=&-t\sum_{n=1}^{N_S-1}(c^{\dg}_{n+1}\tau_zc_{n}+{\rm h.c.})\nn\\ &&+\sum_{n=1}^{N_S}c^{\dg}_{n}\big[-\mu \tau_z
+\De\cos{(\f{\phi}{2})}\tau_x-\De\sin{(\f{\phi}{2})}\tau_y \big]c_{n}, \nn \\
H_D&=& \ep_ac^{\dg}_{0a}\tau_zc_{0a} + \ep_bc^{\dg}_{0b}\tau_zc_{0b}-t_d (c^{\dg}_{0a}\tau_zc_{0b}+{\rm h.c.})\nn\\
H_{TA}&=&-t'(c^{\dg}_{-1}\tau_zc_{0a}+c^{\dg}_{1}\tau_zc_{0b}+{\rm h.c.})\nn\\
H_{TB}&=&-t'(c^{\dg}_{-1}\tau_zc_{0a}+c^{\dg}_{1}\tau_zc_{0a}+{\rm h.c.})~.~\label{eq:ham}
\eea
Here, $H_L$ and $H_R$ are the Hamiltonians for the superconductors on left side and right side of the dot respectively. 
Each superconductor has $N_S$-sites and is modeled by meanfield BdG Hamiltonian. The left~(right) superconductor has
a superconducting phase $\phi/2$~($-\phi/2$), maintaining a phase difference of $\phi$ between the two. 
$c_n=[c_{n,\ua},-c_{n,\da},c^{\dg}_{n,\ua},c^{\dg}_{n,\da}]^T$, where $c_{n,\si}$ is annihilation operator at site $n$
with spin $\si$. $\tau_{x,y,z}$ are the Pauli matrices acting in the 
particle-hole sector. The quantum dot has two sites labeled by $0a$ and $0b$. The two sites have onsite energies
$\ep_{a/b}$ and are connected by a hopping amplitude $t_d$. In configuration-A, the dot site $0a$ is connected to the 
left superconductor and the dot site $0b$ is connected to the right superconductor, while in configuration-B, both 
the superconductors are connected to the dot site $0a$ as can be seen from Eq.~\eqref{eq:ham} and Fig.~\ref{fig-schematic}.

Charge current is not conserved in the superconductors as the Hamiltonians $H_{L/R}$ do not commute with charge operator.
Physically, the Josephson current that flows in the superconductor comes from the electron reservoir (which is not 
explicitly shown in the Hamiltonian) connected to the 
superconductor to maintain its chemical potential. But the charge is conserved in the quantum dot and the charge current
between the superconductor and quantum dot is the Josephson current. The eigenfunctions of the Hamiltonian Eq.~\eqref{eq:ham}
are four-spinors at each site having the form $[\psi_{i,e,n,\ua},\psi_{i,e,n,\da},\psi_{i,h,n,\ua},\psi_{i,h,n,\da}]^T$, where
$i$ is the index denoting different eigenfunctions, the index $e/h$ denotes electron/hole component and $n$ is the site index. 
The eigenspectrum is centered around zero energy. All states below zero energy are filled and all states above zero energy
are empty. The Josephson current calculated at the bond $(-1,0a)$ is given by the formula:
\beq
I(\phi)=\f{2et'}{\hbar}\sum_{E_i<0}\sum_{p=e,h}\sum_{\si=\ua,\da}{\rm Im}[\psi^*_{i,p,-1,\si}\psi_{i,p,0a,\si}] 
\label{eq:current}
\eeq
\section{Results and analysis}
\begin{figure}
 \includegraphics[width=4cm]{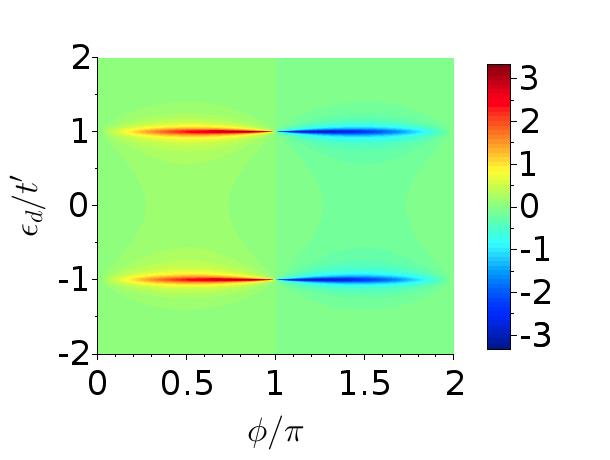}
 \includegraphics[width=4cm]{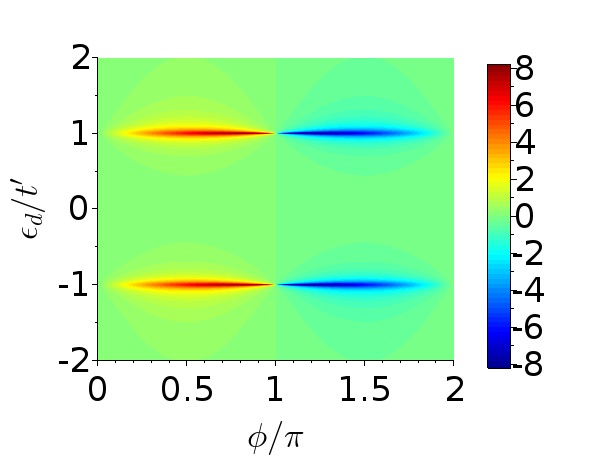}
 \caption{Left panel: $I(\phi)\times10^2$ plotted in units of $et/\hbar$ for configuration~A. 
 Right panel: $I(\phi)\times10^3$ plotted in units of $et/\hbar$ for configuration~B. Parameters chosen for both the 
 configurations:
 $\ep_a=\ep_b=\ep_d$, $N_S=10$, $\De=0.1t$, $t'=t_d=0.1t$ and $\mu=0$.}
 \label{fig-43_e0e1}
\end{figure}

\begin{figure}
 \includegraphics[width=9cm]{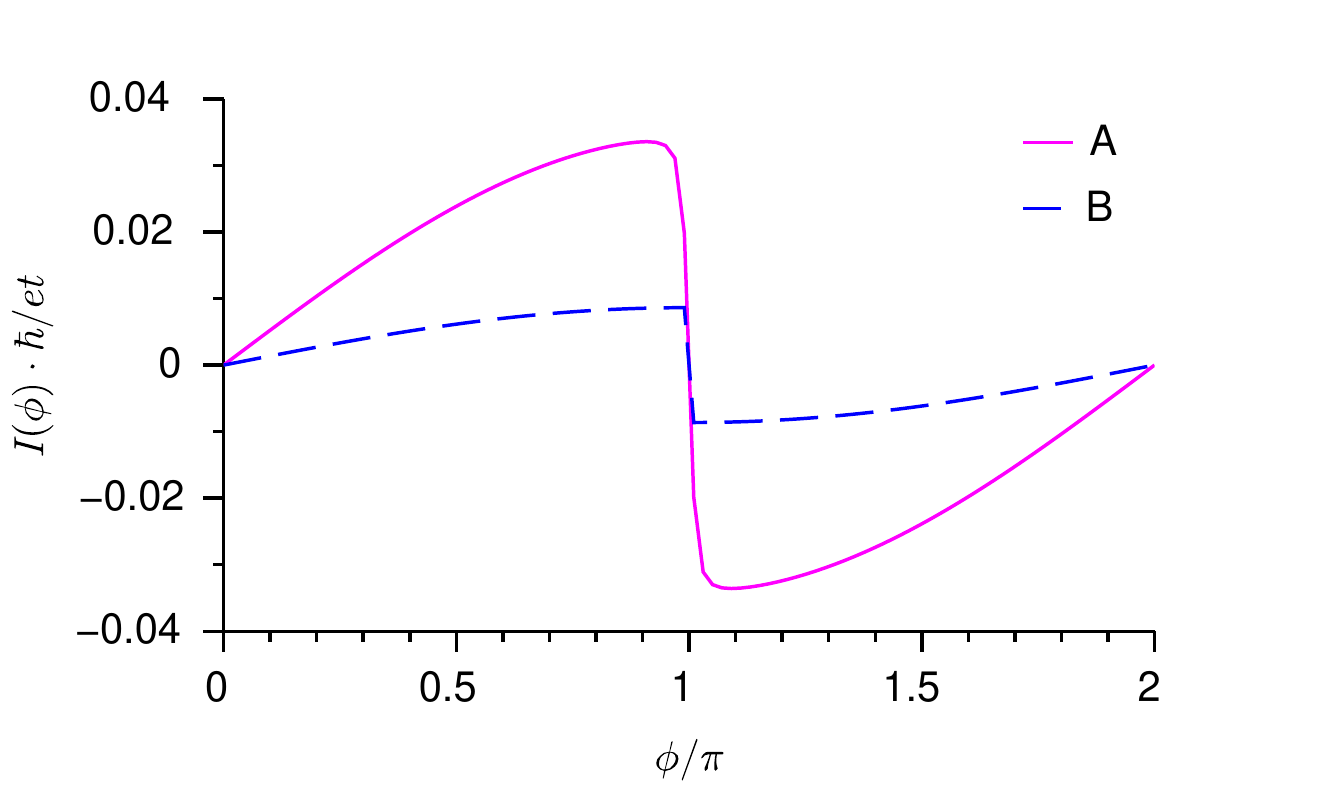}
 \caption{Current phase relation for $\ep_d=t_d=0.1t$. Parameters are same as in Fig.~\ref{fig-43_e0e1}. Different lines
 show the plot for the configurations A and B as shown by the legend.}~\label{fig-cpr-43}
\end{figure}

For demonstrating our primary results, we choose the superconductors of size $N_S=10$. We maintain a superconducting
pair potential of magnitude $\De=0.1t$, where $t$ is the hopping strength in the superconductor. We would like the 
quantum dot to be weakly coupled to the superconductors. So, we choose $t'=0.1t$. The hopping between the two sites
within the quantum dot and the chemical potential in the superconductor take values $t_d=0.1t$ and $\mu=0$ respectively.
First we take the onsite potentials on two sites of the dot to be equal ($\ep_a=\ep_b=\ep_d$) and plot the Josephson
current $I(\phi)$ as a function of the phase difference $\phi$ and $\ep_d$ for the two configurations in Fig~\ref{fig-43_e0e1}.
The plots look similar for the two configurations except that the peak in Josephson current takes a value of about 
$3\times10^{-2}et/\hbar$ for configuration~A and a value of about $8\times10^{-3}et/\hbar$ for configuration~B. The peaks 
in Josephson current occur at $\ep_d=\pm t_d$. This is because, the isolated dot hosts zero energy states when 
$\ep_a=\ep_b=\ep_d=\pm t_d$. Hence the peaks in $I(\phi)$ as a function of $\ep_d$ is a resonance effect.
The plot of $I(\phi)$ versus $\phi$ for $\ep_d=t_d$ is shown in Fig.~\ref{fig-cpr-43} for the 
configurations A~and~B.

\begin{figure}
 \includegraphics[width=4cm]{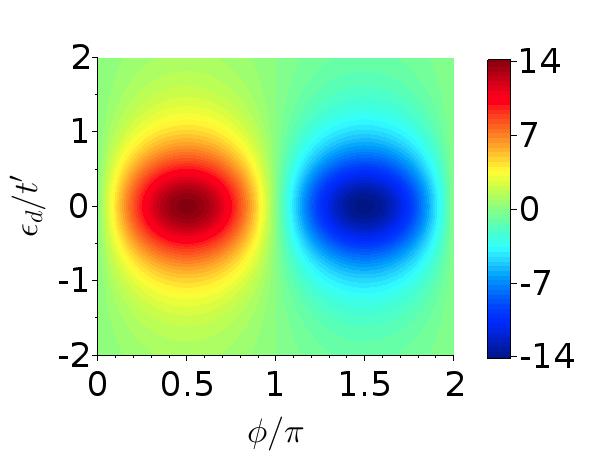}
 \includegraphics[width=4cm]{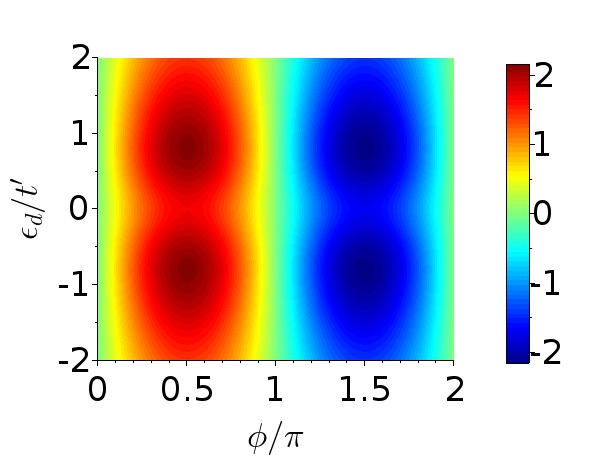}
 \caption{Left panel: $I(\phi)\times10^4$ plotted in units of $et/\hbar$ for configuration~A. 
 Right panel: $I(\phi)\times10^4$ plotted in units of $et/\hbar$ for configuration~B. Parameters chosen for both 
 the configurations:
 $\ep_a=-\ep_b=-\ep_d$, $N_S=10$, $\De=0.1t$, $t'=t_d=0.1t$ and $\mu=0$.}
 \label{fig-43_e0-e1}
\end{figure}

\begin{figure}
 \includegraphics[width=9cm]{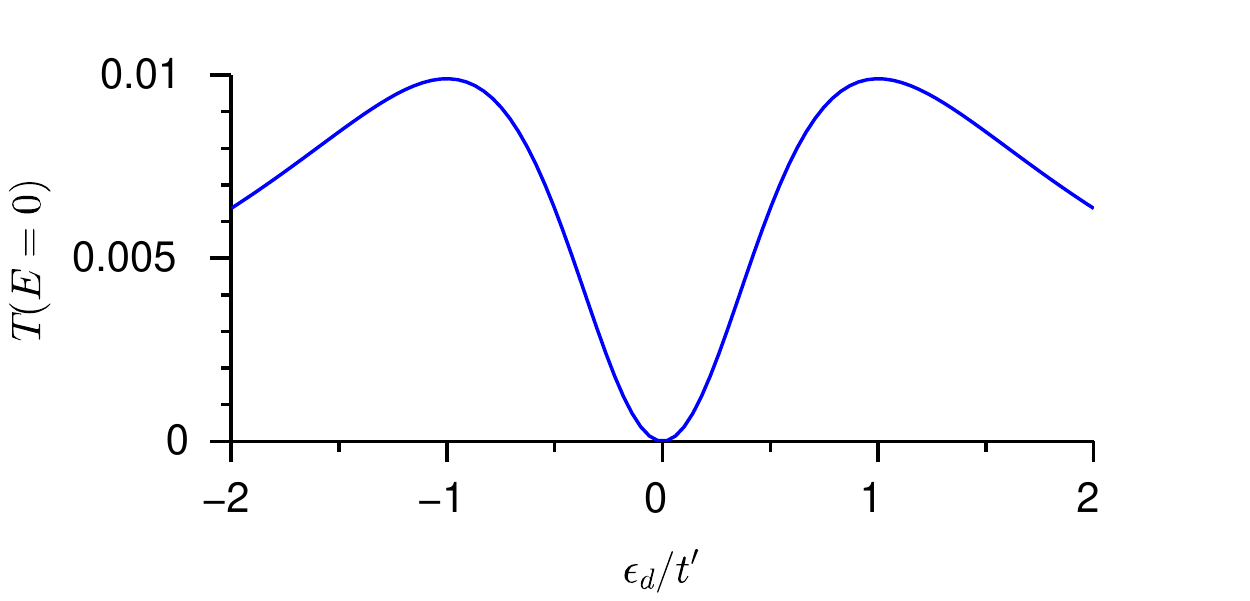}
 \caption{Zero energy transmission probability of the quantum dot coupled to normal metal leads versus $\ep_d$ (where
 $\ep_d=-\ep_a=\ep_b$) given by eq.~\eqref{eq:Tvsed}. Parameters: $t'=t_d=0.1t$.}~\label{fig-7}
\end{figure}

Now, we tweak the two configurations by choosing the onsite potentials on the two sites of the dot to be opposite in sign, 
but equal in magnitude (i.e., $\ep_a=-\ep_b=-\ep_d$). The Josephson currents for this are plotted in Fig.~\ref{fig-43_e0-e1}. 
We see a striking contrast between results for configuration~A and configuration~B. The maximum variation of Josephson current
happens for $\ep_d=0$ for configuration~A, while the same happens for $\ep_d=\pm t_d$ for configuration~B. Further, the 
peaks in the Josephson current versus $\ep_d$ are not sharp as in the case $\ep_a=\ep_b=\ep_d$. Furthermore, the peaks in 
the Josephson current $I(\phi)$ versus $\phi$ occur at $\phi=\pi/2$ as can be seen in Fig.~\ref{fig-43_e0e1}, which is 
very much unlike the case $\ep_a=\ep_b=\ep_d$ where the peak occurs close to $\phi=\pi$ (see Fig.~\ref{fig-cpr-43}). 
In the case $\ep_a=-\ep_b=-\ep_d$, the energy levels of the isolated dot are at $\pm\sqrt{\ep_d^2+t_d^2}$. So, as $|\ep_d|$ 
increases from zero, the dot energy levels move far away from the Fermi energy. Hence, the Josephson current has peaks when
$\ep_d=0$, for configuration~A though the magnitude of the peak current is order of magnitude smaller than that for the case 
$\ep_a=\ep_b=\ep_d$. The result for configuration~B is however puzzling. The puzzle can be resolved if we look at zero-energy
transmission probability of the dot attached to normal metal leads on either sides in configuration~B as a function of $\ep_d$
when $\ep_a=-\ep_b=-\ep_d$. The zero energy transmission probability is given by the expression 
\beq 
T(E=0) = \f{4t'^4\ep_d^2}{4t'^4\ep_d^2+t^2(\ep_d^2+t_d^2)^2}.~\label{eq:Tvsed}
\eeq
In arriving at the above expression, the leads on either sides of the dot are taken to be semi-infinite one dimensional
lattice with chemical potential $\mu=0$ and hopping amplitude $t$. When $T(E=0)$ is plotted versus $\ep_d$ for the same 
parameters as in Fig.~\ref{fig-43_e0-e1}, it shows not-so-sharp peaks at $\ep_d=\pm t_d$ as can be seen in Fig.~\ref{fig-7}.
These peaks in $T(E=0)$ at $\ep_d=\pm t_d$ imply a maximum variation of $I(\phi)$ at $\ep_d=\pm t_d$ in Fig.~\ref{fig-43_e0-e1}.
We numerically find that all our results on the Josephson current remain qualitatively
same in the limit of large sized superconductors (i.e., as $1/N_S$ becomes smaller).

\section{Summary}
To summarize, we have studied a lattice model of two superconductors maintained at a phase difference coupled to each other 
through a quantum dot which has two energy levels for each  spin. The two superconductors can be coupled to each other through
the dot in two geometrical ways which we term as configurations. We first study the case when onsite energies of both the
energy levels of the dot are equal. We examine at the dependence of the Josephson current as a function of the onsite energy 
of the dot and the phase difference. We find resonance effect in Josephson current is exhibited  in both the configurations and the
results in the two configurations are very similar, except for the values of the current. We then tweak the two configurations
by applying onsite energies on the two sites of the dot that are equal in magnitude and opposite in sign. We find that the 
two configurations give different results for the Josephson current. Hence, we show that it is possible to distinguish the two
configurations by choosing the onsite energy profile appropriately. We understand our results in terms of resonant transmission
effect and the  expression for transmission probability of the dot coupled to normal metal leads. 
Our results are within the reach of current experiments where onsite energies of the two sites of the dot can be controlled by 
applying separate gate voltages to the two sites~\cite{saldana18}.
\acknowledgements
The author thanks DST-INSPIRE Faculty Award (Faculty Reg. No.~:~IFA17-PH190)
for financial support. 

\bibliography{refqdjj}
\end{document}